\documentclass[12pt]{article}
\pdfoutput=1

\usepackage{amssymb,amsmath,mathrsfs,epstopdf,slashed,color}
\usepackage{ifpdf}
\ifpdf
\usepackage{graphicx}
\usepackage{hyperref}
\usepackage{cite}
\else
\usepackage[dvipdfmx]{graphicx}
\usepackage[dvipdfmx]{hyperref}
\usepackage[numbers]{natbib}
\usepackage{hypernat}
\fi

\allowdisplaybreaks[1]

\makeatletter

\@addtoreset{equation}{section}
\makeatother

\newcommand{\Vol}{\mathcal{V}}

\DeclareMathOperator*{\re}{Re \,}

\title{\bf Accidental K\"ahler Moduli Inflation}
\author{Anshuman Maharana, Yoske Sumitomo, Markus Rummel}

\begin{document}

\begin{titlepage}

\setcounter{page}{0}
  
\begin{flushright}
 \small
 \normalsize
 KEK-TH-1811
\end{flushright}

\vskip 3cm
\begin{center}

{\Large \textbf{Accidental K\"ahler Moduli Inflation}}

\vskip 2cm
  
{\large Anshuman Maharana${}^{1}$, Markus Rummel${}^{2}$ and Yoske Sumitomo${}^{3}$, }
 
 \vskip 0.6cm

${}^1$Harish Chandra Research Institute,Chhatnag Road, Jhunsi, Allahabad, UP 211019, India\\
${}^2$Rudolph Peierls Centre for Theoretical Physics, University of Oxford,\\ 1 Keble Road, Oxford, OX1 3NP, United Kingdom\\
${}^3$ High Energy Accelerator Research Organization, KEK,\\ 1-1 Oho, Tsukuba, Ibaraki 305-0801 Japan\\

 \vskip 0.4cm

Email: \href{mailto:anshumanmaharana@hri.res, markus.rummel@physics.ox.ac.uk, sumitomo@post.kek.jp}{anshumanmaharana at hri.res, markus.rummel at physics.ox.ac.uk,\\ sumitomo at post.kek.jp}

\vskip 1.0cm
  
\abstract{\normalsize
We study a model of accidental inflation in type IIB string theory where inflation occurs near the inflection point of a small K\"ahler modulus.
A racetrack structure helps to alleviate the known concern that string-loop corrections may spoil K\"ahler Moduli Inflation unless having a significant suppression via the string coupling or a special brane setup.
Also, the hierarchy of gauge group ranks required for the separation between moduli stabilization and inflationary dynamics is relaxed.
The relaxation becomes more significant when we use the recently proposed D-term generated racetrack model.
}
  
\vspace{1cm}
\begin{flushleft}
 \today
\end{flushleft}
 
\end{center}
\end{titlepage}

\setcounter{page}{1}
\setcounter{footnote}{0}

\tableofcontents

\parskip=5pt

\section{Introduction}

Recent observations in cosmology strongly support the mechanism of inflation \cite{Hinshaw:2012aka,Ade:2015lrj,Planck:2015xua}.
During the inflationary epoch, there is an approximately exponential growth of the scale factor, and the quantum fluctuations of the early universe seed the temperature fluctuations in the cosmic microwave background we observe today.
The Planck collaboration measured the value of the spectral index of primordial curvature perturbations precisely, suggesting the behavior of weakly broken scale invariance which is typical in models of inflation \cite{Ade:2015lrj,Planck:2015xua}.
The next experimental goal is to measure the tensor amplitude which is related to the scale of inflation and the displacement of the inflaton in field space.
There have been many attempts to measure the tensor-to-scalar ratio, and the upper bound from the combined data is currently given by $r_{0.002} < 0.11$ \cite{Ade:2015lrj,Planck:2015xua,Ade:2015tva}.
There is much hope that this upper bound will be improved and we will detect a signal for tensor modes soon as significant efforts in the observational front in the near future are being spent towards the detection of primordial B-mode signals (see e.g. \cite{Creminelli:2015oda} and \cite{Matsumura:2013aja}).

Inflation is a high-energy phenomena. The  energy density during inflation is related to the tensor to scalar ratio by $V^{1/4} \sim {\cal O}(0.01) r^{1/4} M_P$ when the COBE normalization is applied.
Even for small values of $r$, the energy scale is not very far away from Planck scale.
Furthermore, the slow-roll conditions make inflation ultra-violet sensitive. This makes it essential to embed inflationary models in an ultra violet complete theory.
String theory is the leading candidate for providing us with a theory of 
quantum gravity. It provides the arena necessary for consistent inflationary model building.
There have been a large amount of works on inflation in string theory (see reviews \cite{Silverstein:2013wua,Quevedo:2014xia,Baumann:2014nda}).
A particularly attractive setting are IIB flux compactifications \cite{Giddings:2001yu,Dasgupta:1999ss}, where moduli stabilization is well developed.

In this paper, we develop a model of inflection point inflation proposed in \cite{Holman:1984yj,Lyth:1998xn} with a K\"ahler modulus in a IIB flux compactification playing the role of the inflaton.
Inflation should begin in a small region around the flat enough inflection point by some accident, and hence is sometimes called \textit{accidental inflation} \cite{Linde:2007jn}.
Accidental inflation has been studied in several contexts in string theory.
Since string  compactifications give rise to complicated potentials, one can expect that an appropriate inflection point which leads to prolonged inflation (avoiding the $\eta$ problem of \cite{Kachru:2003sx}) exists in the vast moduli space.
Familiar examples include D-brane inflation which regards the inflaton field as the location of a mobile D-brane in a warped throat \cite{Dvali:1998pa,Alexander:2001ks,Dvali:2001fw,Burgess:2001fx,Choudhury:2003vr,Kachru:2003sx}.
In the warped D-brane inflation, the compactification and moduli stabilization generate significant contributions to the inflaton potential \cite{Baumann:2006th,Baumann:2008kq,Baumann:2010sx}, and hence we may have an appropriate inflection point where the slow-roll conditions are met \cite{Baumann:2007np,Krause:2007jk,Baumann:2007ah,Agarwal:2011wm,McAllister:2012am} (see also \cite{Chen:2009nk}).
There are, on the other hand, some attempts realizing accidental inflation by an F-term potential in type IIB, in which inflation ends at the KKLT-based de-Sitter minimum ($dS$) \cite{Linde:2007jn} and at the K\"ahler Uplift $dS$ \cite{Ben-Dayan:2013fva} (see also \cite{BlancoPillado:2012cb}). 
String-loop corrections have been used to realize an inflection point potential along the overall volume direction \cite{Conlon:2008cj}. 
In this paper, we will use two types of racetrack models to realize accidental inflation: the ordinary racetrack with multiple non-perturbative effects for a single modulus in the superpotential and the recently proposed D-term generated racetrack model \cite{Rummel:2014raa}, both of which are based on the Large Volume Scenario (LVS) for moduli stabilization \cite{Balasubramanian:2005zx}.

The simplest model of inflation in the LVS involving a K\"ahler modulus is {\it K\"ahler Moduli Inflation } \cite{Conlon:2005jm}.
In this model, the inflation direction is the real part of a K\"ahler modulus which rolls down towards the LVS minimum.
One of the advantages of this model is that we know how  inflation ends so that the resultant reheating mechanism may be understood easily once the matter sector is given.
However, in this model there is the concern that string-loop corrections to supergravity may spoil the inflationary dynamics (see e.g. \cite{Cicoli:2008gp,Burgess:2013sla}).
The leading order string-loop correction can be analyzed explicitly when D7-branes wrap the inflationary cycle.
Similar effects can also be potentially generated by euclidean D3-branes \cite{Baumann:2014nda}.
The basic reason for the problem is the following.
Inflation occurs at a large value of  the K\"ahler modulus to realize a sufficiently flat potential.
Since the term depending on the inflaton field is exponentially suppressed, string-loop corrections that are suppressed by a power law function of the inflaton, can dominate over the inflationary potential and spoil its flatness.
To get a successful model, one has to embed the model in a  compactification in which the loop corrections are tuned small by a small string coupling value or absent by a special brane setup \cite{Cicoli:2008gp,Burgess:2013sla}.

Although the string-loop corrections may also be present, the concern of spoiling the flatness of the inflationary potential is alleviated in our accidental inflation model. The accidental point is located near the minimum, suitable for prolonged inflation, given an appropriate set of coefficients in the racetrack model.
Around the accidental point, the racetrack terms are tuned such that the slow-roll conditions are satisfied by a cancellation between these terms.
Owing to this cancellation, the resultant exponential suppression and hence overall size of each racetrack term becomes modest.
Once there appears a contribution of string-loop corrections, it can be made easily the same order or less than each of the racetrack term with a reasonable value of the string coupling.
The corrections can easily be absorbed by a slight shift of some coefficients of the racetrack potential, while this is difficult in the original K\"ahler Moduli Inflation model unless having a significant suppression via the string coupling or a special brane setup prohibiting the corrections themselves.  
The coefficients of the racetrack terms depend on the complex structure moduli, which in turn are determined by flux quanta.
Therefore the slight shift of coefficients, corresponding to slightly different flux quanta, would be conceivable in a landscape of flux compactifications.

Furthermore, the hierarchy of gauge group ranks $N$ in non-perturbative terms ($a_{\rm inf}/a_{\rm stab} = N_{\rm stab}/ N_{\rm inf}$), that is required between the stabilization and inflation potentials in K\"ahler Moduli Inflation ($a_{\rm inf}/a_{\rm stab} \sim {\cal O}(10)$) \cite{Conlon:2005jm}, is relaxed in the racetrack models to be $a_{\rm inf}/a_{\rm stab} \sim {\cal O} (1)$.
Notably, the effect of this relaxation of hierarchy is more significant in the case of the D-term generated racetrack model.
It is clear that the ratio of the gauge group ranks $a_{\rm inf}/a_{\rm stab} \sim {\cal O} (1)$ is more easily realized generically in string compactifications and introduces less fields in the effective field theory as smaller gauge group ranks $N$ are sufficient.

This paper is organized as follows.
In Section \ref{sec:accid-kaher-moduli}, we illustrate the features of the Accidental K\"ahler Moduli Inflation in the D-term generated racetrack model, starting with the review of original K\"ahler Moduli Inflation.
In Section \ref{sec:ordin-racetr-model}, the superpotential racetrack model is analyzed and compared with the results of the D-term generated racetrack model.
We relegate some of the technical details to Appendix \ref{sec:slow-roll-parameters}.

\section{Accidental K\"ahler Moduli Inflation \label{sec:accid-kaher-moduli}}

In this section, we will present our model of accidental inflation with a K\"ahler modulus as the inflaton. 
The potential will have a racetrack structure generated by a D-term constraint, which was proposed in \cite{Rummel:2014raa}, and will give rise to an accidental inflection point appearing near the uplifted LVS minimum.
This scenario relaxes the concern of K\"ahler Moduli Inflation that the string-loop corrections to the potential spoil the flatness of the F-term inflation potential.
We first start with a review of the D-term generated racetrack model and K\"ahler Moduli Inflation.

\subsection{D-term generated racetrack\label{sec:d-term-generated}}

We start from a model in type IIB string theory with several K\"ahler moduli, defined by
\begin{equation}
 \begin{split}
  &K = -2\ln \left(\Vol + {\xi \over 2}\right), \quad \Vol = (T_1 + \bar{T}_1)^{3/2} - \sum_{i=2}^{5} (T_i + \bar{T}_i)^{3/2},\\
  &W = W_0 + \sum_{i=2}^{5} A_i e^{-a_i T_i},
 \end{split}
 \label{akmi-Kahler and superpotential}
\end{equation}
and the F-term scalar potential $V_F = e^K (|D_I W|^2 - 3 |W|^2)$.
For simplicity, we have omitted the coefficients in the volume, given by the triple intersections of the Calabi-Yau manifold $X_3$, which do not affect our result crucially. The parameter $\xi \propto - \chi g_s^{-3/2}$ is the leading $\alpha'$-correction \cite{Becker:2002nn} with the Euler number $\chi$ of $X_3$.
As we are interested in the parameter region of the LVS \cite{Balasubramanian:2005zx}, any non-perturbative term generated on $T_1$ is negligibly small.
The K\"ahler moduli fields are complex holomorphic fields, which we write as $T_I = \tau_I + i \theta_I$ conventionally.

Recently, in the context of uplifting LVS vacua to de Sitter, a new type of racetrack structure was proposed by using a D-term constraint \cite{Rummel:2014raa}.
When we have magnetized D7-branes wrapping the divisor $D_D$ in the Calabi-Yau, the D-term potential is given by \cite{Haack:2006cy}
\begin{equation}
 \begin{split}
  V_D = {1 \over \re (f_D)} \left(\sum_j c_{Dj} \hat{K}_j \varphi_j - \xi_D \right)^2,
 \end{split}
\end{equation}
with the Fayet-Illiopoulos (FI) term and the gauge kinetic function:
\begin{equation}
 \begin{split}
  &\xi_D = {1\over \Vol} \int_{X_3} D_D \wedge J \wedge \mathcal{F}_D = {1 \over \Vol} q_{D_j} t_j,\\
  &\re (f_D) = {1\over 2} \int_{D_D} J \wedge J - {1\over 2 g_s} \int_{D_D} \mathcal{F}_D \wedge \mathcal{F}_D,
   \end{split}
\end{equation}
where $J = t_i D_i$ is the K\"ahler form on $X_3$ and $q_{D_j} = \tilde{f}_D^k \kappa_{Djk}$ is the anomalous $U(1)$ charge of the modulus $T_j$ induced by the magnetic flux $\mathcal{F}_D = \tilde{f}_D^k D_k$ on $D_D$. $\kappa_{Djk}$ are the triple intersection numbers of $X_3$.
$\varphi_j$ are matter fields charged under the diagonal $U(1)$ of a stack of D7-branes with charges $c_{Dj}$.

Given a choice of divisors $D_i$ and magnetized fluxes $\mathcal{F}_{D_i}$, we may have the D-term potentials
\begin{equation}
 V_{D_4} \propto {1\over \re (f_{D_4})} {1\over \Vol^2} \left( \sqrt{\tau_4} - \sqrt{\tau_3} \right)^2,\\
 V_{D_5} \propto {1\over \re (f_{D_5})} {1\over \Vol^2} \left( \sqrt{\tau_5} - \sqrt{\tau_3} \right)^2,
\end{equation}
where we have assumed that the matter fields are stabilized at either $\varphi_j = 0$ or $\sum c_{Dj} \hat{K}_j \phi_j = 0$ for simplicity.
Since we are interested in the LVS F-term potential where the leading terms appear at ${\cal O} (\Vol^{-3})$, we enforce a D-term constraint of the form $V_{D_4} = V_{D_5} = 0$. Since these appear at  ${\cal O} (\Vol^{-2})$ this means that integrating out the heavy fields corresponds to a constraint
\begin{equation}
 \tau_s \equiv \tau_3 =  \tau_4 =  \tau_5.
  \label{D-term constraint}
\end{equation}
Note that the imaginary parts of the moduli fields associated with the D-term potentials are eaten by massive $U(1)$ gauge bosons with a mass of the string scale through the St\"uckelberg mechanism, owing to the topological coupling of the two-cycle supporting the magnetic flux.

In order to write down the effective potential, we remove the redundancy of the parameters for the dynamics. We redefine
\begin{equation}
 c_i = {A_i \over W_0}, \quad \alpha = {a_3 \over a_2}, \quad \beta = {a_4 \over a_2}, \quad \gamma = {a_5 \over a_2}, \quad \xi_x = a_2^{3/2} \xi,\label{redefinefields}
\end{equation}
and also
\begin{equation}
 \Vol_x = a_2^{3/2} \Vol = a_2^{3/2}  2\sqrt{2} \left(\tau_1^{3/2} - \tau_2^{3/2} - 3 \tau_s^{3/2} \right), \quad x_i = a_2 \tau_i, \quad y_i = a_2 \theta_i.
\end{equation}
Then the F-term scalar potential in the LVS region after imposing the D-term constraint (\ref{D-term constraint}) becomes
\begin{equation}
 \begin{split}
  {V_F \over W_0^2 a_2^3} \sim& {3 \xi_x \over 4\Vol_x^3} + {4 e^{-x_2} c_2 x_2 \over \Vol_x^2} \cos y_2 + {2\sqrt{2} e^{-2x_2} c_2^2 \sqrt{x_2} \over 3 \Vol_x}\\
  &+ {4 e^{-\alpha x_s} \alpha c_3 x_s \over \Vol_x^2} \cos (\alpha y_s) + {4 e^{-\beta x_s} \beta c_4 x_s \over \Vol_x^2} \cos (\beta y_s)+ {4 e^{-\gamma x_s} \gamma c_5 x_s \over \Vol_x^2} \cos (\gamma y_s)\\
  &+ {2\sqrt{2} e^{-2 \alpha x_s} \alpha^2 c_3^2 \sqrt{x_s} \over 3\Vol_x} + {2\sqrt{2} e^{-2 \beta x_s} \beta^2 c_4^2 \sqrt{x_s} \over 3\Vol_x} + {2\sqrt{2} e^{-2 \gamma x_s} \gamma^2 c_5^2 \sqrt{x_s} \over 3\Vol_x}.
 \end{split}
 \label{effective F-term potential}
\end{equation}
Note that $\theta_1$ is stabilized by a non-perturbative effect on $T_1$, which is sub-dominant via an exponential suppression in the overall volume and hence omitted. 
The D-term generated racetrack potential was originally introduced to uplift the LVS minima to stable $dS$ vacua in \cite{Rummel:2014raa}.
Here, however, we like to use the racetrack structure for realization of inflation instead, and so we introduce an uplift term independently by
\begin{equation}
 V = V_F + V_{\rm up}, \quad V_{\rm up} = \left(W_0^2 a_2^{3}\right)  {d_x \over \Vol_x^{4/3}}.
  \label{effective potential}
\end{equation}
The choice of power is motivated by the anti-brane uplifting \cite{Kachru:2002gs,Kachru:2003aw,Kachru:2003sx}.
Note that since inflation will occur mostly along the $x_s$ direction, the power of the uplift does not make any major difference in the following analysis.
Therefore, we just simply use the above uplift term throughout this paper.

\subsection{Review of K\"ahler Moduli Inflation\label{sec:review-kahler-moduli}}

K\"ahler Moduli Inflation \cite{Conlon:2005jm}  can be considered to be a prototypical
model for inflating with a K\"ahler modulus.
Since the K\"ahler Moduli Inflation can be realized just by the single non-perturbative term for the $x_s$ direction, we set $c_3 = c_4 = 0$ in (\ref{effective potential}).
In the parameter regime that realizes the LVS, the volume $\Vol_x$ and $x_2$ are approximately stabilized while $x_s$ rolls down the inflationary potential to a Minkowski ($dS$) minimum.
If we neglect the $x_s$ dependence at the moment (or correspondingly $c_5 = 0$), using a set of parameters:
\begin{equation}
 \begin{split}
  \xi_x = 70, \quad c_2 = -1, \quad d_x \sim 3.234 \times 10^{-5}
 \end{split}
\end{equation}
the Minkowski vacuum sits at
\begin{equation}
 \Vol_x \sim 1782, \quad x_2 \sim 5.979.
  \label{VEV without xs}
\end{equation}
The axion directions are all stabilized at $y_i = 0$.

Next we turn on a potential for $x_s$ with $c_5 \neq 0$.
To realize the stabilization of $\Vol_x, x_2$ regardless of the value of $x_s$ which is necessary for single field inflation, we need a hierarchy between $a_2$ and $a_5$, or a large $\gamma$ as claimed in \cite{Conlon:2005jm}.\footnote{Note that different triple intersection numbers which we did not take into account, can help to improve the hierarchy \cite{Conlon:2005jm}.
In this paper, as we are interested in making a comparison between the original K\"ahler Moduli Inflation and our accidental model on the same footing. Hence, we just focus on the relative ratio $\gamma$ given reasonable triple intersection numbers.}
This can be understood as follows.
Assuming $x_2, \gamma x_s \gg 1$, we solve the extremal equations for $x_2, x_s$ by
\begin{equation}
 \begin{split}
  x_2 \sim {1\over 2} {\cal W}_0 \left( {4 \Vol_x^2 c_2^2 \over 9}\right), \quad x_s \sim {1\over 2 \gamma} {\cal W}_0 \left( {4 \Vol_x^2 \gamma^3 c_5^2 \over 9}\right),\label{x2xsLambert}
 \end{split}
\end{equation}
where we have used the Lambert W-function ${\cal W}(z)$, satisfying $z={\cal W}(z) e^{{\cal W}(z)}$ and expanded by ${\cal W}_0(z) \sim \ln z$ at very large $z$.
This function has two branches of solutions: ${\cal W}_{-1} \leq -1$ and ${\cal W}_0 \geq -1$.
Plugging this solution back in the extremal equation of $\Vol_x$, we get the condition: 
\begin{equation}
 d_x - {27 {\cal W}_0^{3/2} (4\Vol_x^2 \gamma^3 c_5^2 /9) \over 8 \Vol_x^{5/3} \gamma^{3/2}} \sim  -{27 \xi_x \over 16 \Vol_x^{5/3}} + {27 {\cal W}_0^{3/2} (4 \Vol_x^2 c_2^2/9) \over 8 \Vol_x^{5/3}},
\end{equation}
where we put the contributions from the uplift and $x_s$ dependence on LHS. 
Given the values of parameters, the function of $\Vol_x$ on RHS is bounded from above and positive for the parameters of interest.
The term generated through the stabilization of $x_s$ on LHS appears as a negative quantity, suggesting a larger value of $d_x$ to satisfy the constraint at the minimum.
However, around the inflation era, this term related to $x_s$ (the second term in LHS) is not present since $e^{-\gamma x_s}$ is much smaller than $\Vol_x$ for the slow-roll inflation and we do not satisfy the extremal condition for $x_s$.
So, together with the fact that the values of $\Vol_x, x_2$ are not so different at the beginning of inflation and at the minimum, we see that $d_x$ required at the minimum is too large for successful inflation to satisfy the equality due to the upper bound of the RHS.
This situation can be alleviated by increasing $\gamma$ as the subtraction by the $x_s$ related term can be made smaller accordingly.
Given the set of parameters, we have a certain threshold of $\gamma$ that the condition is satisfied regardless of the value of $x_s$.

For the set of parameters:
\begin{equation}
 \xi_x = 70, \quad c_2 = -1, \quad c_5 = -1, \quad (c_3 = c_4 = 0),
\end{equation}
we found
\begin{equation}
 \gamma = 15.5\,,
\end{equation}
as a near minimal value such that $\Vol_x$ and $x_2$ are stabilized regardless of the value of $x_s$ for successful prolonged inflation.
Realizing a Minkowski minimum requires a slight shift of the uplift parameter $d_x$ in the presence of $x_s$ dependence through $c_5\neq 0$, and sits at
\begin{equation}
 d_x \sim 4.137 \times 10^{-5}, \quad \Vol_x^{({\rm min})} \sim 1545, \quad  x_2^{(\rm min)} \sim 5.851, \quad x_s^{(\rm min)} \sim 0.6227.
  \label{kmi-minimum}
\end{equation}
Note that although we have $x_s < 1$, this does not imply a violation of the approximations as we can still satisfy $\gamma x_s >1$ appearing in the exponent, such that higher order instanton corrections are small, as well as $\tau_s > 1$, such that the supergravity approximation is valid.

\begin{figure}[t]
 \begin{center}
  \includegraphics[width=21em]{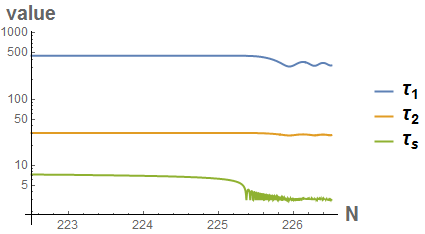}
  \hspace{0em}
  \includegraphics[width=19em]{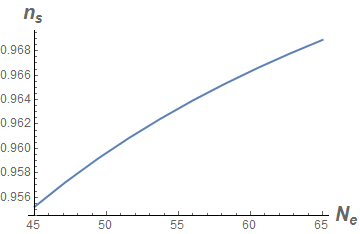}
 \end{center}
 \caption{\footnotesize Inflationary evolution of K\"ahler Moduli Inflation. The inflaton reaches the minimum of the potential at around $N = 226$, while inflation ends at around $N = 224$.}
 \label{fig:kmi-evolution}
\end{figure}

Now we are ready to solve for the inflationary evolution.
We start from the initial point
\begin{equation}
 \Vol_x^{(\rm ini)} \sim 2439, \quad x_2^{(\rm ini)} \sim 6.263, \quad x_s^{(\rm ini)} = 1.8,
  \label{kmi-initial}
\end{equation}
where given the value of $x_s$, the directions $\Vol_x$ and $x_2$ are stabilized.
We impose no initial kick for our analysis.
We also have to specify the overall normalization $W_0^2 a_2^3$ of the potential (\ref{effective potential}) which essentially determines the scale of the power spectrum.
Together with a concrete setup of the rank of gauge groups required to generate the non-perturbative terms, we use
\begin{equation}
 W_0 = {1\over 68}, \quad a_2 = {2 \pi \over 31}, \quad a_5 = {2 \pi \over 2},
\end{equation}
consistent with $\gamma = 15.5$.
This initial point satisfies the slow-roll conditions with
\begin{equation}
 \begin{split}
  \epsilon_{\rm ini} \sim 9.026 \times 10^{-11}, \quad \eta_{\rm ini} \sim - 4.544 \times 10^{-3},
 \end{split}
\end{equation}
where we have used the formula of generalized slow-roll parameters defined in Appendix \ref{sec:slow-roll-parameters}.
For illustration of our parameter choice, we write down the VEV of the moduli fields before the redefinition used in \eqref{redefinefields}.
With the value of $a_2$, we have
\begin{equation}
 \Vol^{(\rm min)} \sim 16930, \quad \tau_2^{(\rm min)} \sim 28.87, \quad \tau_s^{(\rm min)} \sim 3.072
\end{equation}
at the minimum point.

The inflationary trajectory can be solved using the efficient formalism explained in Appendix \ref{sec:slow-roll-parameters} starting from the initial point (\ref{kmi-initial}).
The two relevant plots are shown in Figure \ref{fig:kmi-evolution}.
Inflation ends at $N \sim 224$ where the slow-roll condition breaks down:
\begin{equation}
 \epsilon_{\rm end} \sim 8.503 \times 10^{-6}, \quad \eta_{\rm end} \sim -1.297,
\end{equation}
and further reaches to the minimum point~(\ref{kmi-minimum}) at $N \sim 226$.
Note that $N$ is the e-folding number from the initial point, while $N_e$ is the e-folding number counted backwards from the end of inflation.

It is worth commenting on the the flatness of the potential.
As inflaton is dominantly single field in the $\tau_s$ direction for large values of $\tau_s$, we crudely approximate the slow-roll parameters by
\begin{equation}
 \begin{split}
  &\epsilon \sim K^{\tau_s \tau_s} \left({\partial_{\tau_s} V \over V}\right)^2 \propto \Vol^3 \tau_s^{5/2} e^{-2 a_5 \tau_s},\\
  &|\eta| \sim K^{\tau_s \tau_s} \left({\partial_{\tau_s}^2 V \over V}\right) \propto \Vol^2 \tau_s^{3/2}  e^{-a_5 \tau_s},
 \end{split}
\end{equation}
where we have used $K^{\tau_s \tau_s} \propto \Vol \sqrt{\tau_s}$ and the approximated potential in the inflation era given by $V \sim v_0/\Vol^3 + v_1 \tau_s e^{-a_5 \tau_s}/\Vol^2$ with some constants $v_0, v_1$.
Then the small $|\eta_*| \ll 1$ evaluated at the CMB point implies
\begin{equation}
 \epsilon_* \sim  \eta_*^2 \Vol^{-1} \tau_s^{-1/2} \ll \Vol^{-1} \tau_s^{-1/2}.
\end{equation}
So we see that the value of $\epsilon$ is much smaller than $\eta$ at large volume, resulting in a small value of the tensor to scalar ratio $r$.

We now illustrate some observables at a sample point $N_e = 60$ for a better understanding.
Solving the differential equations \eqref{eq of motion} numerically, we obtain
\begin{equation}
 \epsilon_{60} \sim 1.302 \times 10^{-9}, \quad \eta_{60} \sim -0.01702,
  \label{kmi-slowroll60}
\end{equation}
or correspondingly,
\begin{equation}
 n_s^{(60)}  \sim 0.9660, \quad r^{(60)} \sim 2.08 \times 10^{-8}.
  \label{kmi-nsr}
\end{equation}
Also, the magnitude of the primordial curvature perturbation is estimated by
\begin{equation}
 P_s(k_{60}) \sim 2.165 \times 10^{-9}.
\end{equation}
The cosmological values here nicely agree with the recent results of the Planck collaboration \cite{Ade:2015lrj,Planck:2015xua}.

Let us consider the values of the moduli fields at the $N_e = 60$ point, given by
\begin{equation}
 \Vol_x^{(60)} \sim 2439, \quad x_2^{(60)} \sim 6.263, \quad x_s^{(60)} \sim 1.709,
\end{equation}
and hence the exponential term of the potential important for inflation becomes quite tiny compared to the volume:
\begin{equation}
 e^{-\gamma x_s^{(60)}} \sim e^{-26.50} \sim 3.109 \times 10^{-12} \ll {\Vol^{(60)}}^{-1} \sim 3.742 \times 10^{-5}.
  \label{kmi-exp suppression}
\end{equation}
This large suppression of the term of the inflationary potential is related to the known concern of K\"ahler Moduli Inflation (see e.g. \cite{Cicoli:2008gp,Burgess:2013sla}) as there should exist many corrections to the effective 4D supergravity from string theory.
For instance, if there exists a string loop correction as D7-branes wraps $\tau_s$ (or $\tau_5$), we have a correction to the potential of the form: $\delta V \propto g_s^2 \Vol^{-3} \tau_s^{-1/2}$ \cite{Berg:2007wt,Cicoli:2008va,Cicoli:2007xp}, which may appear significantly larger than the inflation potential term proportional to $\Vol^{-2} e^{-\gamma x_s}$.
Once this term dominates over the exponential potential around the inflation era, the slow-roll parameters turn out to be $\epsilon \propto \Vol \tau_s^{-5/2}, \ \eta \propto \Vol \tau_s^{-2} $ and violate the flatness of the inflationary potential.
Therefore, K\"ahler Moduli Inflation requires a special suppression of these terms via very small $g_s$ to maintain the dominance of F-term potential or a special brane setup which prohibits the existence of the corresponding string loop corrections depending on $\tau_s$.
Note that a similar value of $\gamma x_s^{(60)} \sim 26.5$ is also obtained in the context of Roulette Inflation in \cite{Bond:2006nc}.
In the next section, we like to suggest another possibility of inflation in the direction of a K\"ahler modulus where this concern is alleviated by using an accidental point for inflation.

We have analyzed the potential with the kinetic term obtained from (\ref{akmi-Kahler and superpotential}) imposing the D-term constraint (\ref{D-term constraint}) to make a better comparison with the analysis in the next section.
This kinetic term is slightly different from that of the just three moduli model with $\Vol = (2\tau_1)^{3/2} - (2 \tau_2)^{3/2} - (2 \tau_3)^{3/2}$, although the effective potential is exactly the same.
We have checked that there is no significant difference in the dynamics especially in the observables $n_s, r$ except for ${\cal O} (1)$ factors of $r$ and the e-folding from the beginning to the end of inflation.

\subsection{Accidental K\"ahler Moduli Inflation\label{sec:accid-kahl-moduli}}

In this section, we will explore the possibility that inflation in the $\tau_s$ direction occurs at a point closer to the minimum relative to K\"ahler Moduli Inflation such that the concern about dangerous loop corrections of $\tau_s$ to the dynamics of inflation is relaxed.
To realize successful inflation near the minimum in our setup, one has to tune the parameters in order to achieve the desired flatness of the potential.
Also, the initial condition have to be such that inflation starts close to this tuned flat point.
This inflationary scenario is thus called {\it accidental inflation} as this happens accidentally in the sense of a parametric coincidence.
An accidental model was proposed in \cite{Holman:1984yj}, and accidental inflation in string theory is further studied in \cite{Linde:2007jn,Ben-Dayan:2013fva}.
Here we will illustrate how accidental inflation occurs in the context of K\"ahler Moduli Inflation, especially using a racetrack structure.

\begin{figure}[t]
 \begin{center}
  \includegraphics[width=20em]{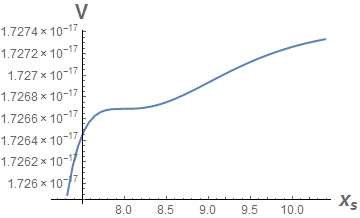}
 \end{center}
 \caption{\footnotesize The accidental point for inflation in Accidental K\"ahler Moduli Inflation. $\Vol_x, x_2$ are approximately stabilized during inflation.}
 \label{fig:akmi-inflection}
\end{figure}

We use the potential defined in (\ref{effective potential}), but with non-zero $c_3, c_4$ to realize the accidental point.
To make a fair comparison, some of the parameters are set to be the same values as in the previous section:
\begin{equation}
 \xi_x = 70, \quad c_2 = -1, \quad c_5 = -1.
\end{equation}
We do not use the large number of the relative ratios of gauge group ranks in the exponent:
\begin{equation}
 \alpha = 2, \quad \beta = 2.1, \quad \gamma = 2.2\,.
\end{equation}
Then the desired accidental point appears when we chose
\begin{equation}
 c_3 = -0.2457004, \quad c_4 = 0.9889189\,.\label{c3c4}
\end{equation}
Note that the number of digits we display in \eqref{c3c4} represents the necessary tuning to realize a flat potential that supports prolonged inflation. The $c_i$ parameters depend on the VEVs of the complex structure moduli and the dilaton which in turn are determined by flux quanta. The huge landscape of possible flux vacua leads to the expectation that these parameters can indeed be tuned to high accuracy.

The minimum of the potential is located at Minkowski with a suitable choice of uplift parameter given by
\begin{equation}
 d_x \sim 3.248 \times 10^{-5}\,.
\end{equation}
Then the minimum sits at
\begin{equation}
 \Vol_x^{(\rm min)} \sim 1775, \quad x_2^{(\rm min)} \sim 5.976, \quad x_s^{(\rm min)} \sim 4.855,
\end{equation}
where all the eigenvalues of the Hessian are defined positive, such that the moduli are stabilized.

The resultant accidental point is illustrated in Figure \ref{fig:akmi-inflection}, where the other moduli are stabilized at
\begin{equation}
 \Vol_x^{(\rm acc)} \sim 1785, \quad x_2^{(\rm acc)} \sim 5.981, \quad x_s^{(\rm acc)} \sim 7.989.
\end{equation}
Using the general formula of the slow-roll parameters presented in Appendix \ref{sec:slow-roll-parameters}, we have
\begin{equation}
 \epsilon_{\rm acc} \sim 2.097 \times 10^{-11}, \quad \eta_{\rm acc} \sim 7.642 \times 10^{-6}.
\end{equation}
The accidental point satisfies the condition $\partial_{x_s}^2 V = 0$ by the tuning of $c_3$ and $c_4$ accordingly. However, $\eta_{\rm acc}$ is not exactly zero due to the presence of off-diagonal entries of both K\"ahler metric and Hessian together with the fact that the accidental point does not satisfy the extremal condition $\partial_{x_s} V = 0$. The value of $\epsilon_{\rm acc}$ is small enough to have prolonged inflation.

Interestingly, the values of moduli fields at the accidental point are not so different from not only those at the minimum, but also those in (\ref{VEV without xs}).
Although we construct the model with small $\alpha, \beta, \gamma$, this essentially means that the $x_s$ direction turns out to be mostly independent of the other directions owing to the racetrack structure.
This feature plays an important role for realizing the accidental point near the minimum.

In the presence of a racetrack potential, one may wonder if the axionic directions are stabilized accordingly, as $c_4 > 0$ contributes as a potential instability at $y_s=0$.
Given the set of parameters, we have
\begin{equation}
 \partial_{y_2}^2  V|_{\rm min} \sim 5.305 \times 10^{-14}, \quad \partial_{y_s}^2 V|_{\rm min} \sim 3.732 \times 10^{-16}\,,
\end{equation}
at the minimum point, while
\begin{equation}
 \partial_{y_2}^2  V|_{\rm acc} \sim 5.218 \times 10^{-14}, \quad \partial_{y_s}^2 V|_{\rm acc} \sim 1.531 \times 10^{-21}\,,
\end{equation}
at the accidental point.
Hence the axionic directions are safely stabilized at $y_i =0$ from the beginning to the end of inflation.

\begin{figure}[t]
 \begin{center}
  \includegraphics[width=21em]{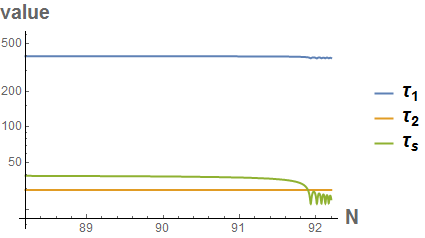}
  \hspace{0em}
  \includegraphics[width=19em]{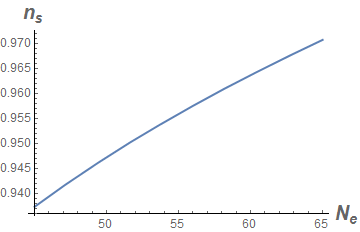}
 \end{center}
 \caption{\footnotesize Inflationary evolution of Accidental K\"ahler Moduli Inflation. The inflaton reaches the minimum of potential at around $N = 92$, while inflation ends at around $N = 90$.}
 \label{fig:akmi-evolution}
\end{figure}

Now we are ready to solve for the evolution of inflation.
We use the accidental point as the initial point.
To match the magnitude of the power spectrum with its observed value, we have to specify the overall coefficient of the potential (\ref{effective potential}). This results in the choice
\begin{equation}
 W_0 = {1\over 55}, \quad a_2 = {2\pi \over 31},
\end{equation}
where we used the same value of $a_2$ as in the previous section.
It may be worth rewriting the VEV at the minimum in terms of $\Vol, \tau_2, \tau_s$ by
\begin{equation}
 \Vol^{(\rm min)} \sim 19450, \quad \tau_2^{(\rm min)} \sim 29.48, \quad \tau_s^{(\rm min)} \sim 23. 95.
\end{equation}
Using the efficient way of solving field evolutions described in Appendix \ref{sec:slow-roll-parameters}, the inflaton rolls down the potential as illustrated in Figure \ref{fig:akmi-evolution}.
The slow-roll condition breaks down at $N\sim 90$ where
\begin{equation}
 \epsilon_{\rm end} \sim 7.283 \times 10^{-6}, \quad \eta_{\rm end} \sim -1.225,
\end{equation}
and the inflaton starts oscillating around the minimum at $N\sim 92$.
Note that we use $N$ for the e-folding from the initial point, while $N_e$ for the e-folding from the end of inflation.

At $N_e = 60$ as a sample point, the slow-roll parameters become
\begin{equation}
 \epsilon_{60} \sim 3.356 \times 10^{-11}, \quad \eta_{60} \sim - 0.01847,
\end{equation}
suggesting
\begin{equation}
 n_s^{(60)}  \sim 0.9631, \quad r^{(60)} \sim 5.369 \times 10^{-10}.
\end{equation}
The values here are not too much different from those in previous section (\ref{kmi-slowroll60}), although the tensor to scalar ratio $r$ is slightly smaller.
The magnitude of the primordial curvature perturbation is given by
\begin{equation}
 P(k_{60}) \sim 2.202 \times 10^{-9}.
\end{equation}
Thus the values of the observables here again nicely agree with the recent data published by the Planck collaboration \cite{Ade:2015lrj,Planck:2015xua}.

\begin{figure}[t]
 \begin{center}
  \includegraphics[width=20em]{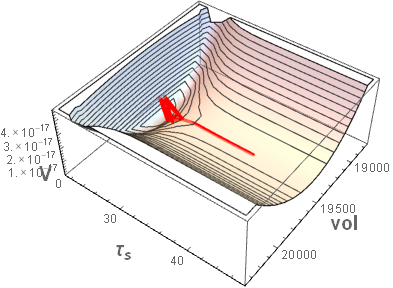}
 \end{center}
 \caption{\footnotesize The potential of Accidental K\"ahler Moduli Inflation at fixed $\tau_2$. The red line describes the trajectory of inflation.}
 \label{fig:akmi-trajectory}
\end{figure}

The form of potential as well as the inflationary trajectory in $\Vol, \tau_s$-space is described in Figure \ref{fig:akmi-trajectory}.
Here we have fixed the $\tau_2$ direction as it is approximately constant during inflation.

Finally, let us evaluate the values of the moduli fields at $N_e = 60$.
These are given by
\begin{equation}
 \Vol_x^{(\rm 60)} \sim 1785, \quad x_2^{(\rm 60)} \sim 5.981, \quad x_s^{(\rm 60)} \sim 7.983.
\end{equation}
Then the exponential suppression in the potential is estimated by
\begin{equation}
 e^{-\alpha x_s^{(60)}} \sim e^{-15.97} \sim 1.165 \times 10^{-7} < {\Vol^{(\rm 60)}}^{-1} \sim 5.111 \times 10^{-5}.
  \label{akmi-suppression}
\end{equation}
When we compare this quantity with the value (\ref{kmi-exp suppression}) of K\"ahler Moduli Inflation, we see the significant alleviation of the suppression term while keeping the successful prolonged inflation.
This alleviation is important.
From the estimation above, we see that each term of the inflationary potential depending on $\tau_s$, for instance $\Vol^{-2} e^{-\alpha x_s}$ is larger or comparable to the string-loop correction of the form $\delta V \sim g_s^2 \Vol^{-3} \tau_s^{-1/2}$ with a reasonable value of the string coupling.
Therefore we conclude that a slight shift of the moduli fields or of the coefficients can absorb the change induced by the corrections as they appear at the same order or less.
This is a feature of Accidental K\"ahler Moduli Inflation improved by the racetrack structure, that is not achieved in the case of K\"ahler Moduli Inflation where the correction may appear as the dominant term without a significant suppression by the string coupling or a special brane setup.

It is worth commenting also that this mild suppression is achieved with more realistic ratios of the gauge group ranks $\alpha, \beta, \gamma \sim {\cal O} (1)$, unlike the situation where a large hierarchy in the exponent $\gamma \sim {\cal O} (10)$ or a larger number of small cycles is required in the original K\"ahler Moduli Inflation.
Also, as a by-product, we could take the larger value $x_s^{(\rm 60)} \sim 7.983$ or $\tau_s \sim 39.39$ as the starting point for the prolonged inflation.
This larger value helps a bit to avoid the string-loop correction $\delta V \propto  g_s^2 {\cal V}^{-3} \tau_s^{-1/2}$ so that the required suppression on $g_s$ is alleviated.

One may wonder if the accidental structure of the potential persist in the presence of sub-leading terms in the overall volume in the F-term potential away from the LVS approximation.
Although these terms are suppressed by the large volume, it might affect our analysis as the leading potential of ${\cal O}(\Vol^{-3})$ includes another cancellation due to the racetrack structure for slow-roll inflation at the accidental point.
Analyzing the full F-term potential, we see that actually these sub-leading terms do not spoil the existence of both minimum and accidental point.
At the accidental point where the additional cancellation works, the structure of the potential consists mostly of $e^{-\alpha x_s}/\Vol_x, e^{-\beta x_s}/\Vol_x, e^{-\gamma x_s}/\Vol_x$ terms, and the resultant potential changes little in the presence of sub-leading terms.
This small change can be easily absorbed  by a slight shift of the coefficients $c_3, c_4$ so that a successful phase of prolonged inflation is realized.

\section{Ordinary racetrack model\label{sec:ordin-racetr-model}}

So far we have studied the racetrack model generated through the D-term constraint.
In this section, we study the standard racetrack model with triple non-perturbative terms in the superpotential. 
To be more precise, we consider the following model:
\begin{equation}
 \begin{split}
  &K = -2\ln \left(\Vol + \xi/2\right), \quad
  \Vol = \left(T_1 + \bar{T}_1\right)^{3/2} - \left(T_2 + \bar{T}_2\right)^{3/2} - \left(T_3 + \bar{T}_3\right)^{3/2},\\
  &W = W_0 + A_2 e^{- a_2 T_2} + A_3 e^{- a_3 T_3} + B^{(1)}_3 e^{-b_3^{(1)} T_3} + B^{(2)}_3 e^{-b_3^{(2)} T_3},
 \end{split}
\end{equation}
where we have used a slightly different volume form compared to the previous section.
In this model, the effective potential is of the same form as in the previous section except for the cross-terms of a racetrack structure as we will see below.
We assume that there is no D-term constraint for this model.

We first redefine the parameters to remove the redundancy in the dynamics by
\begin{equation}
 \begin{split}
  &c_{2} = {A_2 \over W_0}, \quad c_a= {A_3 \over W_0}, \quad c_{b} = {B_3^{(1)} \over W_0} \quad c_{c} = {B_3^{(2)} \over W_0}, \\
  &\alpha = {a_3 \over a_2}, \quad \beta = {b_3^{(1)} \over a_2}, \quad \gamma = {b_3^{(2)}\over a_2}, \quad \xi_x = a_2^{3/2} \xi,
 \end{split}
\end{equation}
and
\begin{equation}
 \Vol_x = a_2^{3/2} \Vol, \quad x_i = a_2 \tau_i, \quad y_i = a_2 \theta_i.
\end{equation}
Using the redefined parameters above, the F-term potential $V_F = e^{K} (|DW|^2 - 3 |W|^2)$ together with the uplift term of (\ref{effective potential}) becomes
\begin{equation}
 \begin{split}
  {V \over W_0^2 a_2^3} \sim& {d_x \over \Vol_x^{4/3}} + {3 \xi_x \over 4 \Vol_x} + {4 e^{-x_2} c_2 x_2 \over \Vol_x^2} \cos y_2 + {2\sqrt{2} e^{-2 x_2} c_2^2 \sqrt{x_2} \over 3 \Vol_x}\\
  &+{4 e^{-\alpha x_3} \alpha c_a x_3 \over \Vol_x^2} \cos (\alpha y_3) +{4 e^{-\beta x_3} \beta c_b x_3 \over \Vol_x^2} \cos (\beta y_3) +{4 e^{-\gamma x_3} \gamma c_c x_3 \over \Vol_x^2} \cos (\gamma y_3)\\
  &+ {2\sqrt{2} e^{-2 \alpha x_3} \alpha^2 c_a^2 \sqrt{x_3} \over 3 \Vol_x} + {2\sqrt{2} e^{-2 \beta x_3} \beta^2 c_b^2 \sqrt{x_3} \over 3 \Vol_x} + {2\sqrt{2} e^{-2 \gamma x_3} \gamma^2 c_c^2 \sqrt{x_3} \over 3 \Vol_x}\\
  &+ {4\sqrt{2} e^{-(\alpha+\beta)x_3} \alpha \beta c_a c_b \sqrt{x_3} \over 3\Vol_x} \cos \left((\alpha-\beta) y_3 \right) + {4\sqrt{2} e^{-(\beta+\gamma)x_3} \beta\gamma c_b c_c \sqrt{x_3} \over 3\Vol_x} \cos \left((\beta - \gamma) y_3 \right)\\
  &+ {4\sqrt{2} e^{-(\gamma + \alpha)x_3} \gamma \alpha  c_c c_a \sqrt{x_3} \over 3\Vol_x} \cos \left((\gamma - \alpha) y_3 \right).
 \end{split}
 \label{rkmi potential}
\end{equation}
The last three cross-terms are typically present in case of a superpotential racetrack model as they show up at ${\cal O}(\Vol^{-3})$.
Again the uplift term motivated by an anti-brane uplift is just a choice and the result does not change crucially if we use a different power of the volume dependence, corresponding to a different uplift mechanism.

Now we solve the dynamics.
The parameters responsible for the size of the volume are chosen as
\begin{equation}
 \xi_x = 70, \quad c_2 = -1, \quad c_c = -1,
\end{equation}
such that we can make a fair comparison to the results in the previous section.
Unlike the previous situation in the D-term generated racetrack model, it is difficult to realize the stabilization of $\Vol_x, x_2$ during a phase of successful prolonged inflation in the $x_3$ direction unless having hierarchical values of the coefficients $\alpha, \beta, \gamma$.
So we set
\begin{equation}
 \alpha = 7, \quad \beta = 7.1, \quad \gamma = 7.2,
\end{equation}
to illustrate a successful example.
With the choice of parameters above, the accidental point is achieved with the remaining parameters being
\begin{equation}
 c_a = -0.73971879, \quad c_b = 1.71976665,
\end{equation}
and the axion directions are stabilized at $y_i = 0$ as before.
We are now ready to determine the Minkowski minimum which demand a parameter value
\begin{equation}
 d_x \sim 4.086 \times 10^{-5}.
\end{equation}
Then the minimum point sits at
\begin{equation}
 \Vol_x^{(\rm min)} \sim 1570, \quad x_2^{(\rm min)} \sim 5.866, \quad x_3^{(\rm min)} \sim 0.6080,
\end{equation}
where all the eigenvalues of the Hessian are positive.

\begin{figure}[t]
 \begin{center}
  \includegraphics[width=20em]{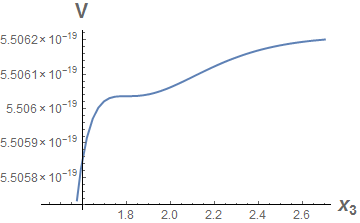}
 \end{center}
 \caption{\footnotesize The accidental point for inflation in the ordinary racetrack model. $\Vol_x, x_2$ are approximately stabilized during inflation.}
 \label{fig:rkmi-inflection}
\end{figure}

When we use a smaller choice of $\alpha, \beta, \gamma$, the required value for the uplift $d_x$ becomes slightly larger.
In fact, this slightly larger value of the uplift parameter is dangerous as this causes a destabilization of the volume direction at larger values of $x_3$ where inflation occurs.
This is not the case in the D-term generated racetrack model as we have illustrated in Section \ref{sec:accid-kahl-moduli}, while we have this situation in the ordinary racetrack model.
This could be due to the existence of cross-terms appearing in the last three terms of (\ref{rkmi potential}).
We set two of the coefficients $c_a, c_b, c_c$ to negative values to realize moduli stabilization, suggesting negative entries of the cross-terms at the minimum.
To compensate the negative contribution of cross-terms, we have to increase the uplift parameter $d_x$.
Hence, 
we have to start with slightly larger values of $\alpha, \beta, \gamma$ to maintain the hierarchy that the cross-term contribution does not violate the stability of $\Vol_x, x_2$.

With the suitable choice of parameters for our purpose, the accidental point is located at
\begin{equation}
 \Vol_x^{(\rm acc)} \sim 2272, \quad x_2^{(\rm acc)} \sim 6.199, \quad x_3^{(\rm acc)} \sim 1.800,
\end{equation}
as illustrated in Figure \ref{fig:rkmi-inflection} where $\Vol_x, x_2$ are approximately stabilized.
When we use the general expression of slow-roll parameters in Appendix \ref{sec:slow-roll-parameters}, we estimate
\begin{equation}
 \epsilon_{\rm acc} \sim 6.332 \times 10^{-13}, \quad \eta_{\rm acc} \sim 6.028 \times 10^{-6}.
\end{equation}
Again, the value of $\eta_{\rm acc}$ is not exactly zero due to the off-diagonal entries and the non-zero value of the first derivative $\partial_{x_3} V \neq 0$ although satisfying $\partial_{x_3}^2 V = 0$.
We will use this accidental point as the initial point for inflation.

Before proceeding, we comment on the stability of the axionic directions as it is not obvious in the racetrack model.
At the extremum $y_i = 0$, there is no mixture of the Hessians between the real and imaginary modes of the moduli fields.
The dominant entries of the Hessian of the axionic directions becomes
\begin{equation}
 \partial_{y_2}^2 V|_{\rm min} \sim 4.246 \times 10^{-17}, \quad \partial_{y_3}^2 V|_{\rm min} \sim 1.126 \times 10^{-16}
\end{equation}
at the minimum point and
\begin{equation}
 \partial_{y_2}^2 V|_{\rm acc} \sim 1.536 \times 10^{-17}. \quad \partial_{y_3}^2 V|_{\rm acc} \sim 1.083 \times 10^{-23}
\end{equation}
at the accidental point.
So the axionic directions are indeed stabilized during the entire inflationary evolution.

\begin{figure}[t]
 \begin{center}
  \includegraphics[width=21em]{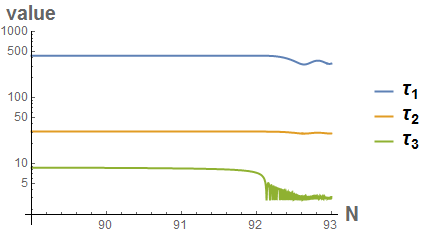}
  \hspace{0em}
  \includegraphics[width=19em]{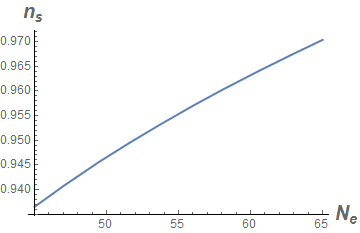}
 \end{center}
 \caption{\footnotesize Inflationary evolution in the ordinary racetrack model. The inflaton reaches the minimum of the potential at around $N = 93$, while inflation ends at around $N = 90$.}
 \label{fig:rkmi-evolution}
\end{figure}

Next we discuss the inflationary solution to the field equations \eqref{eq of motion}.
To fix the overall scale of the potential, we choose
\begin{equation}
 W_0 = {1\over 2300}, \quad a_2 = {2\pi \over 31},
\end{equation}
where the value of $a_2$ is chosen identical to the previous section to make a fair comparison.
With this choice, the values of moduli fields at the minimum are rewritten by
\begin{equation}
 \Vol^{(\rm min)} \sim 17210, \quad \tau_2^{(\rm min)} \sim 29.94, \quad \tau_3^{(\rm min)} \sim 3.000. 
\end{equation}
Solving by the efficient method described in Appendix \ref{sec:slow-roll-parameters}, the slow-roll condition for inflation breaks down at $N \sim 90$ where the slow-roll parameters are estimated as
\begin{equation}
 \epsilon_{\rm end} \sim 2.206 \times 10^{-7}, \quad \eta_{\rm end} \sim -1.123,
\end{equation}
and the inflaton reaches the minimum at $N\sim 93$.
We illustrate the evolution in Figure \ref{fig:rkmi-evolution}, where $N$ is the e-folding from the initial point and $N_e$ is the e-folding from the end of inflation.

Now we estimate some observables at $N_e = 60$ as a sample point.
The slow-roll parameters become
\begin{equation}
 \epsilon_{60} \sim 1.071 \times 10^{-12}, \quad \eta_{60} \sim -0.01874,
\end{equation}
which implies
\begin{equation}
 n_s^{(60)} \sim 0.9625, \quad r^{(60)} \sim 1.714 \times 10^{-11}.
\end{equation}
Then the magnitude of the primordial curvature perturbation is estimated as
\begin{equation}
 P(k_{60}) \sim 2.196 \times 10^{-9}.
\end{equation}
The observables shown here are well in agreement with the recent results of the Planck collaboration \cite{Ade:2015lrj,Planck:2015xua}.

Finally, we present the values of moduli fields for the validity of the approximation.
At $N_e = 60$, the moduli fields have values of
\begin{equation}
 \Vol_x^{(60)} \sim 2272, \quad x_2^{(60)} \sim 6.199, \quad x_3^{(60)} \sim 1.799.
\end{equation}
Using the values above, we get
\begin{equation}
 e^{-\alpha x_3^{(60)}} \sim e^{-12.59} \sim 3.405 \times 10^{-6} < {\Vol^{(60)}}^{-1} \sim 4.017 \times 10^{-5}.
  \label{rkmi-suppression}
\end{equation}
Although we had to choose slightly larger $\alpha, \beta, \gamma$ to maintain an appropriate hierarchy for the stabilization of $\Vol_x, x_2$, the resultant suppression of the inflation potential is quite alleviated compared to K\"ahler Moduli Inflation.
This is due to the fact that we could start from an initial point near the minimum point, owing to the racetrack structure.
Note that the suppression here looks slightly milder than in the case of the D-term generated racetrack model.
This could be because we did not try to find an extreme choice of parameters that relaxes the suppression as much as possible.
Therefore we consider that the suppressions of the ordinary racetrack model and the D-term generated racetrack model are very similar, and conclude that this alleviation of the suppression is a generic feature of the racetrack models with a suitable accidental point.

Note that the accidental structure persists even in the presence of sub-leading terms in the volume in the F-term potential away from the LVS approximation, similar to Section \ref{sec:accid-kahl-moduli}.
Again, the sub-leading terms change the potential little, and a slight shift of the coefficients $c_a, c_b$ can easily absorb the small changes in the potential.
This can be anticipated since the resultant contribution of the inflaton field to the potential (\ref{rkmi-suppression}) is actually comparable with that in the D-term generated racetrack model (\ref{akmi-suppression}), where the potential is dominated by single exponent terms even in the presence of a racetrack cancellation.

\section{Discussion}

We have presented Accidental K\"ahler Moduli Inflation where inflation occurs around an accidental point realized by a racetrack structure. The observable values obtained, i.e., the magnitude of primordial curvature perturbation, spectral index and tensor-to-scalar ratio are within the observational bounds.
The concern that string-loop corrections spoil the flatness of the inflaton potential, which is present in the original K\"ahler Moduli Inflation model, is alleviated in the Accidental K\"ahler Moduli Inflation model.
Inflation occurs at an accidental point near the minimum, avoiding the exponentially large hierarchy of values of the inflationary potential terms between the inflationary era and the minimum. Hence, the correction term can be absorbed in the inflationary potential accordingly.
Similar arguments should apply to other potentially dangerous corrections, making our construction rather robust.

We have presented two types of racetrack models: one uses the D-term generated racetrack structure proposed in \cite{Rummel:2014raa} and the other consists of the ordinary racetrack structure in the superpotential.
Given a set of reasonable parameters in the D-term generated racetrack model, a phase of successful prolonged inflation can be realized by a reasonable ratio of  gauge group ranks $a_{\rm inf}/a_{\rm stab} = \alpha, \beta, \gamma \sim {\cal O}(1)$ in the exponents, while the original K\"ahler Moduli Model demands $a_{\rm inf}/a_{\rm stab} \sim {\cal O}(10)$ to realize a sufficient hierarchy between volume modulus stabilization and inflation.
In fact, the possible rank of the gauge groups generating the non-perturbative effects is constrained from above by the consistency with tadpole cancellation and holomorphicity \cite{Collinucci:2008pf,Cicoli:2011qg,Louis:2012nb}, and hence it may not be easy to achieve such large hierarchies of $\gamma \sim {\cal O}(10)$ in explicit models.
Also, since the D-term generated racetrack model is constructed with a larger number of K\"ahler moduli, the maximal gauge group rank satisfying the consistency conditions is actually larger than that of model starting with smaller number of moduli \cite{Louis:2012nb}. However, the constraint on $\gamma$ also becomes less severe in the original model of K\"ahler Moduli Inflation when the number of small cycles increases.
To summarize, we consider the D-term generated racetrack model for accidental inflation is realizable in the sense of explicit model constructions in string theory.
Note that the ordinary racetrack model requires slightly larger coefficients to maintain the hierarchy, but a phase of prolonged inflation is realized with $a_{\rm inf}/a_{\rm stab} < 10$ and thus is reasonable too.

Although the realization of the accidental point requires the tuning of coefficients and initial conditions, it does not require unnatural hierarchies of the coefficients $c_i = A_i / W_0$.
The coefficient of each racetrack term is determined by the stabilization of the complex structure moduli and dilaton, and hence we expect that the coefficients appear at the same order of magnitude in general, which is actually the case for the models considered in this paper.

The racetrack structure can be realized without unacceptable differences of ratios $\alpha, \beta, \gamma$ in the exponents.
The difference of $(\beta - \alpha) \sim {\cal O}(10^{-1})$ etc. is easily realized by taking slightly different gauge group ranks of D7-branes wrapping the inflationary cycle.
Although we used that each ratio differed by `0.1' as an illustration throughout this paper, the system works even with larger differences.
In addition, for the D-term generated racetrack model, the D-term constraint may be generalized to be $\tau_3 = \delta_4 \cdot \tau_4 = \delta_5 \cdot \tau_5$ with a more complicated configuration of magnetic flux on D7-branes associated with the D-terms, where $\delta_{4}, \delta_{5} \neq 1$ are some rational numbers.
This different choice of D-term constraints affects the ratios $\alpha, \beta, \gamma$ in the exponents, and hence it certainly helps to realize even smaller differences of ratios, resulting in a better racetrack with less tuning of coefficients $c_i$ of each racetrack term.
Another advantage of the D-term generated racetrack model is that we do not need to worry about the detailed physics related to the split of gauge group ranks which would be required to realize the ordinary racetrack terms in the superpotential.


Even though the existence of an inflection point is conceivable in a complicated string moduli space, one may worry about the overshooting problem of initial conditions.
When we start from a higher point along the inflationary direction, more e-folds are achieved since the potential around the inflection point is flat enough.
However, given the values of coefficients ($A_i, W_0$), the resultant spectral index at CMB point differs from the observational value with the different initial condition.
If we were to allow changing the values of the coefficients simultaneously with the initial condition, some region in the parameter space would be allowed. 
Although there is a discussion of this issue in the context of string theory \cite{Itzhaki:2007nk,BlancoPillado:2012cb}, we hope to report more on this issue in the future.

One may worry about isocurvature perturbations which are currently severely constrained by the Planck collaboration \cite{Ade:2015lrj,Planck:2015xua}.
The light mode contributing to isocurvature perturbations is regarded as a massless particle during inflation. In our model this is the axionic partner of the overall volume modulus.
The other fields are heavy enough and hence the adiabatic perturbations are essentially driven by the single inflaton field.
The mass of $\theta_1$ is generated by a non-perturbative effect on $T_1$, that is sub-dominant in LVS and we did not specify in our model.
If the mass scale is large enough to be a dark matter, this axion also contributes to the CDM isocurvature perturbations.
Since the magnitude of CDM isocurvature perturbation crucially depends on the initial mis-alignment angle and the fraction of CDM consisting of axion, we believe that the CDM isocurvature constraint is satisfied (see e.g. Figure 6 of \cite{Kawasaki2008} at $H_{\rm inf} \sim 10^{10} \ {\rm GeV}, \ f_a \sim 10^{16} \ {\rm GeV}$ of our illustrative example).
On the other hand, when the mass scale is negligible even at the minimum of LVS, this lightest axion could play a role of dark radiation.
In this case, although the direct entropy production from axion perturbations would not be subject to the constraint as the axion only has derivative couplings and hence the entropy is diluted away quickly, the dark radiation produced by the decay of the volume modulus may contribute to the neutrino isocurvature density perturbations \cite{Kawasaki2012}.
However, a more conclusive discussion requires more details of the setup including the location of the matter sector in the compact manifold. Hence, we consider this is beyond the scope of the paper.

So far, we have focused on  accidental inflation which occurs near an inflection point. It may be interesting to see what happens when inflation starts near a saddle point instead, as studied in \cite{BlancoPillado:2004ns,BlancoPillado:2006he}.
Although axionic saddle points were used in \cite{BlancoPillado:2004ns,BlancoPillado:2006he}, when inflation occurs along the small modulus direction, two racetrack terms may be able to form a saddle point in the small modulus direction.
However, not only $\epsilon$, but also $\eta$ needs to be suppressed for successful inflation.
To realize this, we may need an additional racetrack term or other contributions to the potential.
Saddle point inflation along the small modulus direction also needs the correct initial conditions that the inflaton does not roll down a runaway direction which would result in decompactification.
We hope to pursue this direction in the future.

\section*{Acknowledgments}
We are grateful for stimulating discussions with Hideo Kodama, Kentaro Tanabe and Alexander Westphal.
We would like to thank the organizers of the KEK Theory Workshop 2015 where this work was initiated.
This work is partially supported by the Grant-in-Aid for Scientific Research (A) (26247042) from the Japan Society for the Promotion of Science (JSPS).

\appendix

\section{Slow-roll parameters and dynamics\label{sec:slow-roll-parameters}}

In this appendix, we denote the details of the inflationary dynamics as well as the definition of the slow-roll parameters.
We follow the efficient way of solving the full set of equations of motions for the inflationary trajectories used in \cite{BlancoPillado:2006he}.

First, the kinetic term is written by
\begin{equation}
 {\cal L}_{\rm kin} = {\partial^2 K \over \partial T_I \bar{T}_J} \partial_\mu T_I \partial^\mu \bar{T}_J = -{1\over 2} g_{ab} \partial_\mu \phi^a \partial^\mu \phi^b,
\end{equation}
where $T_I = T_1, T_2, T_3, T_4, T_5$ and $\phi^a = \tau_1, \tau_2, \tau_s, \theta_2, \theta_s$ in the D-term generated racetrack model of Section \ref{sec:accid-kaher-moduli}.
Note that we have imposed the D-term constraint (\ref{D-term constraint}) in the second equation, and have neglected $\theta_1$.
We use a different coordinate system accordingly in the ordinary racetrack model of Section \ref{sec:ordin-racetr-model}.
Using the non-canonical metric, the generalized slow-roll parameters are defined by
\begin{equation}
 \epsilon = {1\over 2} g^{ab} {\partial_a V \partial_b V \over V^2},
\end{equation}
and the most negative eigenvalue of the matrix
\begin{equation}
 N^a_{\ b} = g^{ac} {\partial_c \partial_b V - \Gamma^d_{cb} \partial_d V \over V}
\end{equation}
gives the other slow-roll parameter $\eta$.
$\Gamma^d_{cb}$ is the connection obtained form $g_{ab} (\phi)$ in the normal fashion.
It may also be convenient to write down the slow-roll parameters in terms of the complex coordinates:
\begin{equation}
 \begin{split}
  &\epsilon = K^{I \bar{J}} {\partial_I V \partial_{\bar{J}} V \over V^2},
  \qquad N^a_{\ b} = \left(\begin{array}{cc}
            N^I_{\ J} & N^I_{\ \bar{J}}\\ N^{\bar{I}}_{\ J} & N^{\bar{I}}_{\ \bar{J}}
                           \end{array}\right),\\
  &N^{I}_{\ J} = K^{I \bar{K}} {\partial_{\bar{K}} \partial_J V \over V}, \quad
  N^{I}_{\ \bar{J}} = K^{I\bar{K}} {\partial_{\bar{J}}\partial_{\bar{K}} V - K^{L\bar{M}} \partial_{\bar{J}} \partial_{\bar{K}} \partial_{L} K \partial_{\bar{M}} V \over V}.
 \end{split}
\end{equation}
Note that $N^{\bar{I}}_{\ \bar{J}}, N^{\bar{I}}_{\ J}$ are complex conjugate of $N^{I}_{\ J}, N^{I}_{\ \bar{J}}$.

The inflationary evolution can be calculated efficiently when we introduce the canonical momenta defined by
\begin{equation}
 \begin{split}
  \pi_a = {\partial {\cal L}_{\rm kin} \over \partial \dot{\phi}^a}.
 \end{split}
\end{equation}
Using these equations, we solve $\dot{\phi}^a$ as functions of $\pi_a$.
Then the equation of motions become
\begin{equation}
 \begin{split}
  {d \phi^a \over d N} =& {1 \over H} \dot{\phi}^a (\pi),\\
  {d \pi_a \over d N} =& -3 \pi_a - {1\over H} {\partial \over \partial \phi^a} \left( V - {\cal L}_{\rm kin} \right),\\
  3 H^2 =& \rho = {\cal L}_{\rm kin} + V,
 \end{split}
 \label{eq of motion}
\end{equation}
where $N$ is the number of e-foldings from the initial point of inflation.
As explained in \cite{BlancoPillado:2006he}, we can avoid computing Christoffel symbols directly in field space owing to the introduction of the canonical momenta $\pi_a$, and hence save some computational cost.

We may approximate the equations of motion (\ref{eq of motion}) under the slow-roll approximation.
When the inflation trajectory is well approximated by a single field and the other fields are stabilized, the equation becomes
\begin{equation}
 3 H^2 {d \phi^{\rm i} \over d N} + g^{ii} \partial_i V = 0,
\end{equation}
where $i$ denotes the direction of inflation, which is $\tau_s$ or $\tau_3$ in the models we consider.
This approximate equation suggests
\begin{equation}
 N_e \sim \int_{\phi^i_{\rm end}}^{\phi^{i}_{\rm CMB}} d\phi^i {V \over g^{ii} \partial_i V},
\end{equation}
where $N_e$ is the e-folding from the end of inflation and we have used  values at the end of inflation $\phi^i_{\rm end}$ and at the CMB point $\phi^{i}_{\rm CMB}$.
In the accidental scenario we study in this paper, all values of $N, N_e$ are estimated by solving the full equations of motion (\ref{eq of motion}), while the equation under the slow-roll approximation gives mostly the same results except for a few small numerical differences.

The power spectrum of adiabatic scalar density perturbations under the slow-roll condition is estimated by
\begin{equation}
 P_s(k) = {1\over 8 \pi^2} {H^4 \over {\cal L}_{\rm kin}} \sim {1 \over 24 \pi^2} {V \over \epsilon},
\end{equation}
which is evaluated at the horizon crossing point $k = a H \sim H e^N$.
Therefore the spectral index becomes
\begin{equation}
 n_s - 1 = {d \ln P_s(k) \over d \ln k} \sim {d \ln P_s (N) \over d N} \sim 2 \eta - 6 \epsilon.
\end{equation}
Finally, the tensor to scalar ratio is given by
\begin{equation}
 r = {P_t(k) \over P_s(k)} \sim 16 \epsilon.
\end{equation}

\bibliographystyle{utphys}
\bibliography{myrefs}

\end{document}